\documentclass[10pt,letterpaper,twocolumn]{article} 

\usepackage{ol2}
\usepackage[draft]{hyperref}
\usepackage{amsmath}

\begin{document}

\twocolumn[ 

\title{Broadband Phase-Noise Suppression in a Yb-Fiber Frequency Comb}


\author{A. Cing\"{o}z,$^{1,*}$ D. C. Yost,$^1$ T. K. Allison,$^1$ A. Ruehl,$^2$ M. E. Fermann,$^2$ I. Hartl$^2$ and J. Ye$^1$}

\address{
$^1$JILA, National Institute of Standards and Technology and University of Colorado, Boulder, Colorado 80309-0440, USA
\\
$^2$IMRA America, Inc., 1044 Woodridge Avenue, Ann Arbor, MI 48105, USA \\
$^*$Corresponding author: acingoz@jila.colorado.edu
}

\begin{abstract}
We report a simple technique to suppress high frequency phase noise of a Yb-based fiber optical frequency comb using an active intensity noise servo. Out-of-loop measurements of the phase noise using an optical heterodyne beat with a continuous wave (cw) laser show suppression of phase noise by $\geq$7~dB out to Fourier frequencies of 100~kHz with a unity-gain crossing of $\sim$700~kHz. These results are enabled by the strong correlation between the intensity and phase noise of the laser. Detailed measurements of intensity and phase noise spectra, as well as transfer functions, reveal that the dominant phase and intensity noise contribution above $\sim$100~kHz is due to amplified spontaneous emission (ASE) or other quantum noise sources.
\end{abstract}


 ] 
\noindent Fiber based frequency combs have received much attention recently~\cite{Fermann2009} because of several key advantages in comparison to solid-state Ti:sapphire frequency combs~\cite{Udem2002,Cundiff2003}. Fiber based combs are less expensive, more robust, and allow for turn-key operation. In addition, while Er-based combs are compatible with fiber-optic components developed for telecommunications, Yb-based similariton combs have achieved low-frequency noise operation and high average power~\cite{Schibli2008,Ruehl2010}.

One of the most important applications of frequency combs has been in frequency metrology where the repetition frequency, $f_{rep}$, and the carrier envelope offset frequency, $f_{0}$, must be phase locked to either optical or radio-frequency (rf) references in order to facilitate comparison of widely disparate optical frequencies. Progress in fiber comb technology has led to Er combs with sub-Hertz linewidths~\cite{Swann2006} and Yb combs with sub-mHz linewidths~\cite{Schibli2008}, suitable for this application. However, for a broad range of more recent applications, such as the coupling of the frequency comb into a Fabry-Perot optical cavity for the purpose of enhanced trace molecular detection~\cite{Thorpe2006}, or the production of intracavity high-harmonic generation (HHG)~\cite{Jones2005,Gohle2005}, it is especially necessary to minimize the broadband phase noise of the comb teeth in the optical domain to optimize coupling and mitigate phase-to-amplitude noise conversion by the cavity.

The most prominent sources of broadband phase noise in dispersion-compensated fiber-based combs are the residual intensity noise (RIN) of the pump diode and the relatively large contribution of ASE. The latter leads to direct pulse-to-pulse timing jitter in addition to residual intensity noise~\cite{Haus1993,Paschotta2004}. Changes in the pump diode output lead to changes in both the gain and pulse parameters of the oscillator, which affect both $f_{rep}$ and $f_0$ via a variety of nonlinear mechanisms~\cite{Newbury2005}. However, in practical terms, the action of pump diode RIN and other noise terms is best described by the fixed-point formalism~\cite{Benkler2005}, where a given perturbation leads to changes in $f_{rep}$ and $f_0$ such that the spacing between comb teeth breathe about a fixed frequency at which the changes due to $f_{rep}$ and $f_0$ cancel. Due to this coupled action, error signals derived from $f_{rep}$ or $f_0$ alone do not directly address the action of the perturbation and are less effective at minimizing the broadband noise in the optical domain compared to error signals derived directly from the RIN.

Study of pump-induced frequency jitter in fiber combs has been reported in Refs.~\cite{McFerran2006,Budunoglu2009}. In Ref.~\cite{McFerran2006}, a portion of this noise was suppressed passively in an Er-based frequency comb by high power operation of the pump diode, which led to a reduction of the RIN of the pump diode and a suppression of the oscillator phase noise up to a Fourier frequency of $\sim$80~kHz.

In this letter, we study the effects of RIN on a Yb-based frequency comb at several pump powers above mode-locking threshold and show that the measured frequency noise power spectral density is correlated with the intensity noise spectrum of the oscillator. Comparison of pump modulation transfer functions and intensity noise spectra reveal that quantum noise sources dominate over pump RIN above $\sim$100~kHz. Hence, in contrast to Ref.~\cite{McFerran2006}, we use an active intensity noise cancelation servo that minimizes the total oscillator intensity noise. Out-of-loop measurements show a phase noise suppression for the comb teeth near the carrier frequency of $\geq$7~dB out to a Fourier frequency of 100~kHz, with a servo unity-gain crossing point of $\sim$700~kHz.

\begin{figure}[t]
\centerline{
\includegraphics[width=8.3cm]{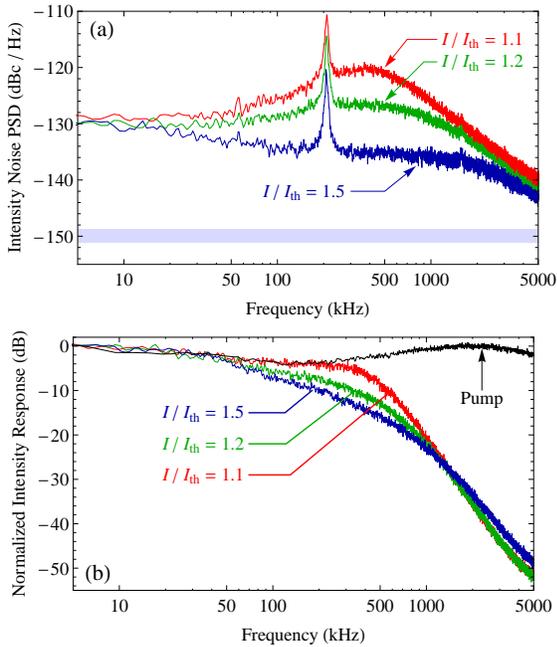}}
\caption{\label{fig:1} (Color online) (a) Intensity noise spectrum of the Yb oscillator at various pump powers normalized to DC signal. The shaded box indicates the shot noise floor of the measurements. (b) Intensity modulation transfer functions for the oscillator and the pump diode.}
\end{figure}

The Yb frequency comb used for this experiment was recently described in Ref.~\cite{Ruehl2010}. This laser produces an average power of 80~W through chirped pulse amplification of the oscillator pulse train. The amplitude-to-frequency noise conversion is dominated by the nonlinearities in the oscillator since linear, saturated amplification is employed, which should not affect the phase noise. The oscillator consists of a Fabry Perot-type Yb-fiber laser. It is mode locked with a subpicosecond lifetime saturable absorber and dispersion compensated by a chirped fiber Bragg grating for operation in the similariton regime. The optical oscillator output spectrum bandwidth is $\sim$45~nm, centered at 1050~nm with $>$100~mW average power. The repetition frequency of the laser is 154~MHz.

The intensity noise of the oscillator was detected using a home-built fiber-coupled photodetector with 40~dB of gain, a bandwidth of 10~MHz, and shot-noise limited performance with $>$20~$\mu$W of optical power. The intensity noise spectra at various currents, $I$, above the mode-locking threshold, $I_{th}$, are shown in Fig.~\ref{fig:1}a. The spectra have been normalized to their respective DC signal levels. The oscillator intensity response shows a relaxation-oscillation peak at $\sim$400~kHz near the mode-locking threshold, which is damped as the pump power is increased. The spike near 200~kHz is most likely technical in nature. In comparison, the measured pump diode RIN level is $\sim$130~dBc/Hz at DC and is flat at few dB out to a Fourier frequency of 5~MHz. The measured decrease in the pump diode RIN between 600~mA of current ($I<I_{th}$) and 1~A ($I\sim1.5$~$I_{th}$) is $\sim$2~dBc. Figure~\ref{fig:1}b shows the response of the oscillator and the pump diode to pump current modulation. These transfer function were obtained by applying white noise to the current modulation input of the pump diode. While the pump diode transfer function is flat out to 5~MHz, limited by the driver electronics, the oscillator response to pump current modulation differs from the intensity noise spectra at high frequencies. For example, near threshold, the intensity noise falls off as 8~dB/octave beyond 1~MHz, while the modulation transfer function falls off as $\sim$15~dB/octave, indicating that noise sources other than the pump RIN are responsible for the observed intensity noise of the oscillator at high frequencies.

\begin{figure}[tb]
\centerline{
\includegraphics[width=8.3cm]{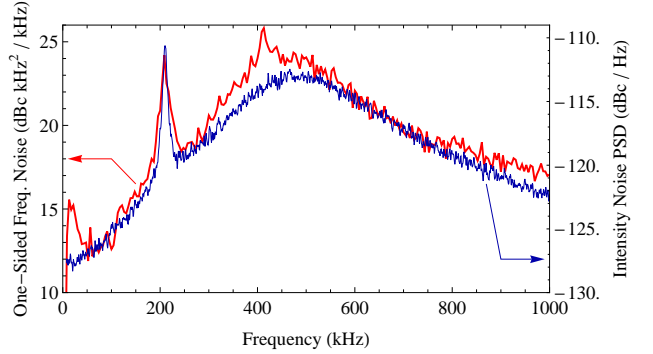}}
\caption{\label{fig:2} (Color online) Comparison of the intensity noise spectrum with the frequency noise of a comb tooth near the optical carrier for $I\approx1.1$~$I_{th}$.}
\end{figure}

Noise sources other than the pump RIN, which are collectively named quantum noise, include ASE, fluctuations in the linear losses in the cavity, and the vacuum fluctuations entering through the output coupler~\cite{Harb1997}. The response of the oscillator intensity to pump and quantum fluctuations can be described by the transfer functions $G_p(f)$ and $G_{qn}(f)$, respectively. The relationship between the two transfer functions is given approximately by $G_{qn}(f)\propto f^2\,G_p(f)$. This can be understood using the semi-classical rate equations where the fluctuations in the pump power change the gain, while the quantum fluctuations act directly on the pulse energy. These simple considerations are in agreement with more rigorous treatments, such as analytic quantum models for cw lasers~\cite{Harb1997} and numerical simulations of mode-locked lasers~\cite{Paschotta2004}. Thus, we compare a simple model of the form $a\,G_p+b\,G_{qn}$ to the measured intensity noise spectra to retrieve the relative contribution of the noise sources. This analysis reveals that the quantum noise begins to dominate over the pump noise at Fourier frequencies of 70, 80, and 180~kHz for pump current conditions of 1.1, 1.2, and 1.5 times $I_{th}$, respectively.

The frequency noise spectrum near the optical carrier frequency was measured by loosely phase locking a comb tooth to a cw 1064-nm non-planar ring oscillator (NPRO) using a piezoelectric actuator that controls the oscillator cavity length. Figure~\ref{fig:2} shows the comparison of the intensity and frequency noise spectra for $I/I_{th}\approx1.1$. The frequency spectrum between 50~kHz and 1~MHz is strongly correlated with the intensity spectrum, indicating that amplitude-to-frequency noise conversion in the oscillator due to both pump and quantum noise dominates other phase/frequency noise processes such as direct timing jitter due to ASE.

\begin{figure}[tb]
\centerline{
\includegraphics[width=8.3cm]{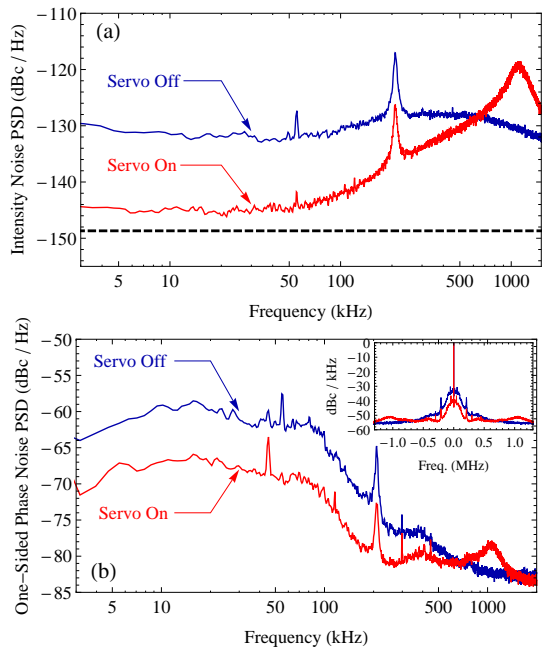}}
\caption{\label{fig:3} (Color online) (a) Intensity noise spectrum of the oscillator with and without the servo action. The dashed line indicates the shot-noise floor. (b) Reduction in the phase noise near the carrier frequency due to the servo. Inset: rf spectrum of the phase-locked optical beatnote.}
\end{figure}

Given the strong correlation between the oscillator intensity and frequency noise, an active intensity noise servo was constructed to reduce the high-frequency phase noise of the oscillator. The error signal for the servo was derived from the oscillator output. The loop filter consisted of a single proportional-integral stage, in addition to two cascaded lead stages in order to combat the 2-pole amplitude roll-off and phase lag introduced by the measured transfer function shown in Fig.~\ref{fig:1}b. The output of the servo was fed back to the current modulation input of the pump diode.

Figure~\ref{fig:3}a shows the action of the noise servo on the in-loop intensity noise of the oscillator at a pump current of $I/I_{th}=1.2$. The noise servo reduces intensity noise by $\sim15$~dB out to a Fourier frequency of $\sim$100~kHz with a unity gain frequency of 700~kHz. The performance of the servo was characterized out of loop using the heterodyne optical beat between the comb and the 1064-nm NPRO. The one-sided phase noise power spectral density with and without the action of the intensity noise servo is shown in Fig.~\ref{fig:3}b, which reveals $\sim$7~dB phase noise suppression out to 100~kHz and once again, a unity gain frequency of 700~kHz. The integrated phase noise from 700~kHz to 2~kHz (below which phase noise due to environmental perturbations dominates over that due to RIN) drops from 0.1~rad$^2$ to 0.027~rad$^2$.

The amount of phase noise suppression depends on the pump current and can be as high 10~dB out to 300~kHz with $I/I_{th}\sim 1$. Nevertheless, for all appropriate current values, the absolute phase noise spectrum and the unity gain frequency achieved with the servo turned on is essentially the same after adjustments to the loop filter.

The noise peaking past the unity gain frequency present in both intensity and phase noise spectra leads to an additional integrated phase noise of 0.008~rad$^2$ above 1~MHz. However, this noise is inconsequential for femtosecond enhancement cavity applications since the response of the cavity rolls off as a single pole above the cavity linewidth. Indeed, the transmission intensity noise spectrum of a high-finesse cavity (linewidth 100~kHz) locked to the comb showed a reduction of 10~dB out to 100~kHz Fourier frequency with the intensity servo on and no detectable increase in higher frequency noise.

In summary, we have demonstrated an active intensity noise servo capable of actuating on the oscillator RIN and suppressing the correlated phase noise with high bandwidth. These results are especially pertinent to the production of vacuum ultraviolet (VUV) frequency combs through the use of intracavity HHG~\cite{Jones2005,Gohle2005} since excess amplitude noise within the enhancement cavity can add phase noise to the generated VUV frequency comb during the highly nonlinear HHG process~\cite{Yost2009}. Moreover, the robust and high bandwidth nature of this noise suppression technique should make it applicable to a variety of fiber-laser frequency stabilization schemes.

We acknowledge helpful discussions with T. R. Schibli. This research is funded by DARPA, NIST, and NSF. A. Cing\"{o}z and T. K. Allison are National Research Council postdoctoral fellows. A. Ruehl acknowledges funding from the A. von Humboldt Foundation (Germany) and is currently at Institute for Lasers, Life and Biophotonics, Vrije Universiteit Amsterdam.


\end{document}